\documentclass[11pt]{article}
\usepackage[utf8]{inputenc}
\usepackage[T1]{fontenc}
\usepackage[margin=1in]{geometry}
\usepackage{graphicx}
\usepackage{amsmath}
\usepackage{url}
\usepackage{placeins}
\usepackage{authblk}
\usepackage{apacite}

\AtBeginDocument{%
  \urlstyle{same}%
}

\title{Dynamic Evolution of Pore-scale Heterogeneity and Transport Conditions Control Mineral Dissolution Regimes}

\author[1,2]{Jinlei Wang}
\author[2,*]{Yongfei Yang}
\author[1]{Martin J. Blunt}
\author[1,*]{Branko Bijeljic}
\affil[1]{Department of Earth Science and Engineering, Imperial College London, London, UK}
\affil[2]{Research Center of Multiphase Flow in Porous Media, School of Petroleum Engineering, China University of Petroleum (East China), Qingdao, China}
\affil[*]{Corresponding authors: \texttt{yangyongfei@upc.edu.cn}; \texttt{b.bijeljic@imperial.ac.uk}}
\date{}

\begin{document}

\maketitle


\begin{abstract}
Mineral dissolution in porous media is classically partitioned into static regimes within the $Pe$--$Da$ plane, but this framework fails to capture the dissolution behavior of structurally complex rocks. Using three-dimensional micro-continuum simulations on micro-CT images of three rock samples spanning a wide range of pore-space heterogeneity, we track the joint evolution of dissolution morphology, velocity distribution, and reaction rate. Our results reveal that initial flow heterogeneity controls accessibility of reactants, thereby controlling the dissolution regime, reshaping them as dynamic trajectories. Channeled dissolution emerges as a simultaneous reorganization of structure and flow, and the resulting permeability-porosity relationship cannot be captured by a single power-law. The effective power-law exponent increases with heterogeneity and changes over time, reaching a maximum of 9.8, 18.0, and 40.9 for the three samples. Consequently, the effective reaction rate falls one to three orders of magnitude below the uniform dissolution prediction, with the suppression scaling with flow heterogeneity due to mass transfer limitations in channeled dissolution.
\end{abstract}

\noindent\textbf{Key Points:}
\begin{itemize}
\item Permeability-porosity dynamic exponent and velocity distributions quantify how heterogeneity and transport conditions control dissolution regimes.
\item Pore-scale flow heterogeneity drives channeled dissolution, breaking the simple power-law scaling of permeability with porosity.
\item Only homogeneous media experience uniform dissolution; reduction in effective reaction rates is more rapid with increasing heterogeneity.
\end{itemize}


\section{Introduction}

Reactive transport in porous media \cite{steefel1996approaches, kang2006lattice}, characterized by the dynamic coupling between fluid flow and mineral dissolution, is a fundamental process in many subsurface natural and industrial systems, including geological CO$_2$ storage \cite{krevor2023subsurface, steefel2013pore}, groundwater remediation \cite{fryar1998hydraulic}, geothermal energy extraction \cite{harlow1972theoretical, liu2025thermal}, oil and gas recovery \cite{liu2017modeling, yang2023recent}, and karst formation \cite{clemens1996combined}. The spatiotemporal evolution of pore-space geometry dictates the macroscopic transport properties \cite{detwiler2003experimental, panga2005two}. Predicting these morphological changes requires an understanding of the complex interplay among fluid dynamics, species transport and geochemical reactions \cite{lee2025effects, menke2015dynamic, li2022effects}.

Mineral dissolution processes are classically categorized into three regimes: face, wormhole and uniform \cite{fredd1998influence, hoefner1988pore}, defined by the competition between mass transfer and reaction rates, typically quantified by the dimensionless Péclet ($Pe$) and Damköhler ($Da$) numbers. Face dissolution arises when reaction is fast relative to mass transfer, wormhole arises when reaction and mass transfer are comparable, and uniform dissolution arises when reaction is slow relative to mass transfer \cite{al2019pore, golfier2002ability, wang2025multi}. Traditional phase diagrams proposed to predict dissolution regimes in the $Pe$-$Da$ plane \cite{golfier2002ability, soulaine2017mineral, szymczak2009wormhole, wang2022transitions, wang2023reactive} have proven valuable for determining the macroscopic dissolution behavior. However, conventional regime boundaries within these diagrams were primarily derived from homogeneous or weakly heterogeneous media, and treat the dissolution regime as a fixed classification rather than a dynamic process that evolves with the pore structure.

Natural rocks are topologically complex porous media with pore-scale heterogeneity. This pore structure dictates preferential flow channels that alter dissolution behavior \cite{al2017reaction, bijeljic2013predictions, steefel2005reactive}. This flow heterogeneity has a significant impact on the dissolution regimes and reaction rates\cite{jung2018physical, kanavas2025flow, szawello2024quantifying, oliveira2020multispecies}. Laboratory experiments and numerical simulations have revealed non-uniform dissolution where uniform dissolution would be predicted using the traditional phase diagram \cite{menke2016reservoir, pereira2016pore, roded2021wormholing, yang2020dynamic}. Channeled dissolution, in which preferential flow paths are widened along their entire length, has been identified as a distinct class in structurally complex rocks \cite{al2018reservoir, menke2023channeling, menke2016reservoir, pereira2016pore}. The macroscopic reaction rate in heterogeneous rocks is typically several orders of magnitude below the intrinsic surface rate, a discrepancy due to incomplete reactant access to the mineral surface once preferential channels emerge \cite{menke2015dynamic, molins2012investigation, deng2018pore, li2008scale}. However, pore-scale heterogeneity and its amplification into flow heterogeneity have not been systematically quantified across dissolution regimes, and the impact of this heterogeneity on the macroscopic effective reaction rate remains poorly understood.

To quantify the impact of heterogeneity and transport conditions on dissolution regimes and reaction rates, we perform 3D micro-continuum simulations on three rock samples (calcite beadpack, Ketton limestone, and Estaillades limestone) spanning a wide range of pore-scale heterogeneity, tracking the joint evolution of dissolution regimes and permeability-porosity relationships. We propose that the flow heterogeneity, quantified by the velocity distributions and the coefficient of variation $CV_v$ of the flow field, organizes the coupling of structure and flow and selects the regime each rock develops, recasting the dissolution regime as a dynamic trajectory rather than a fixed label.

Our objectives are: (i) to establish a methodology for quantifying the impact of heterogeneity on dissolution regimes and their transitions; (ii) to identify what porous media sustains uniform dissolution and what initial flow heterogeneity is required to produce channeled dissolution; (iii) to examine the dependence of the $K$--$\phi$ relationship on heterogeneity and transport conditions; and (iv) to investigate the reduction in the effective reaction rate as a result of mass transfer limitations induced by heterogeneity across dissolution regimes.

\section{Materials and Methods}

\subsection{Sample Characterization}

\begin{figure}[!ht]
 \centering
 \includegraphics[width=\textwidth]{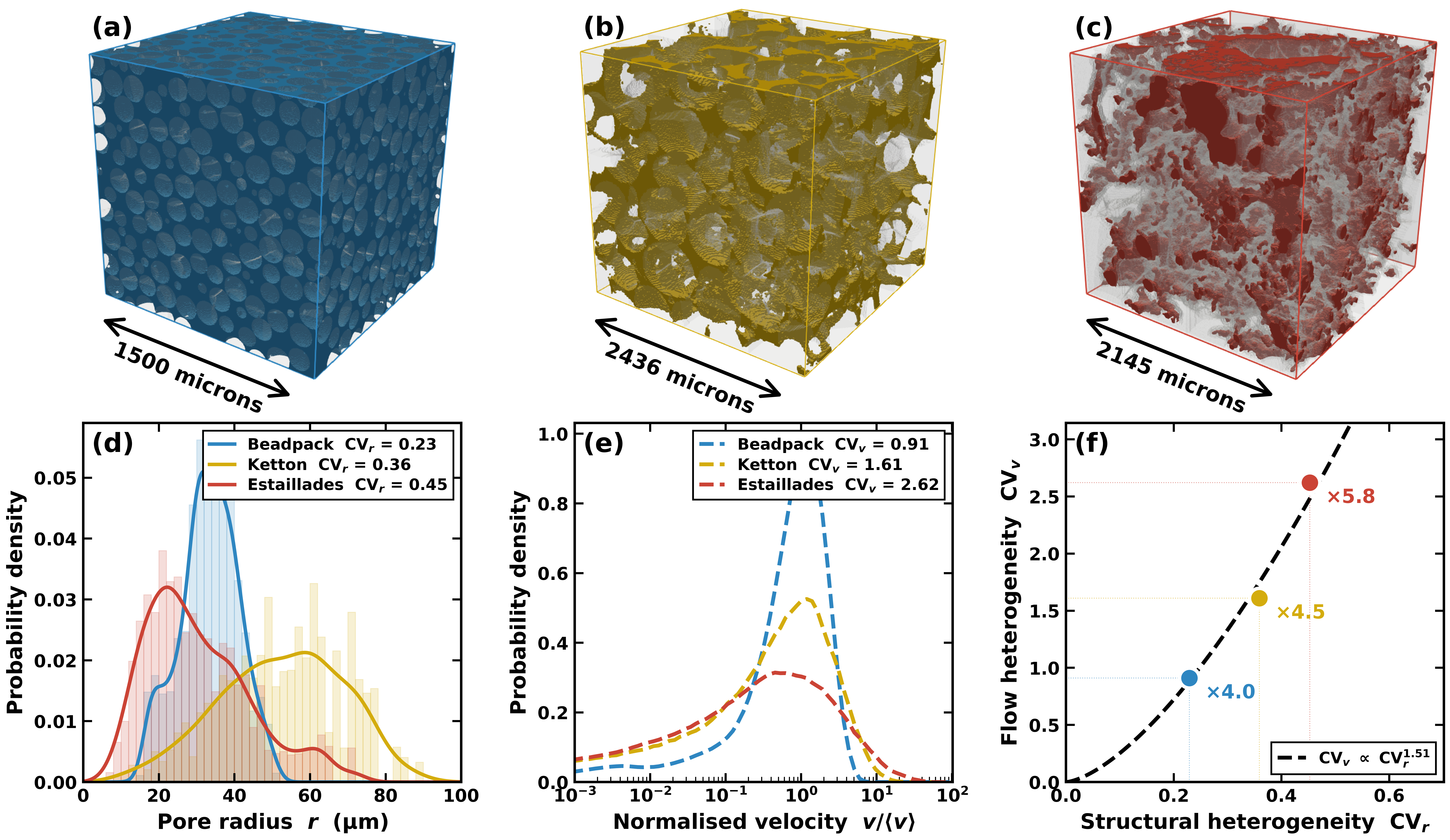}
 \caption{Pore-scale characterization and heterogeneity quantification of the three representative rock samples. (a--c) 3D visualizations of (a) the beadpack, (b) Ketton, and (c) Estaillades, where pore spaces are highlighted in distinct colors (blue, yellow, and red) within the semi-transparent mineral. (d) Probability density functions, PDFs, of the pore radius. (e) PDFs of the normalized velocity. (f) Superlinear scaling between $CV_v$ and $CV_r$, $CV_v \propto CV_r^{1.51}$.}
 \label{fig:fig1}
\end{figure}

To investigate the effect of pore-scale heterogeneity on mineral dissolution regimes, we selected three samples (Figure \ref{fig:fig1}a--c) spanning a wide range of pore topological complexity: beadpack, Ketton limestone and Estaillades limestone \cite{finney1970random, andrew2014pore, ma2026pore}. Their porosities are 0.362, 0.140 and 0.119, and permeabilities are 13.9, 2.7 and 0.5\,D, respectively (Table~S1). The micro-CT image volumes are $500^3$, $400^3$, and $650^3$ voxels with voxel sizes of 3.0, 6.1, and 3.3\,$\mu$m, respectively. The beadpack, a random packing of identical spheres, serves as a homogeneous benchmark with a well connected pore space. We characterize pore structure heterogeneity by the coefficient of variation of the pore-radius probability density function (PDF), $CV_r$ (Figure~\ref{fig:fig1}d), and the flow heterogeneity by the coefficient of variation of the pore-velocity PDF, $CV_v = \sigma/\mu$ (Figure~\ref{fig:fig1}e). $CV_v$ increases from the beadpack (0.91) through Ketton (1.61) to Estaillades (2.62). The two metrics scale super-linearly, $CV_v \propto CV_r^{1.51}$, with amplification factors $CV_v/CV_r$ of 4.0, 4.5, and 5.8 for the beadpack, Ketton and Estaillades, respectively (Figure~\ref{fig:fig1}f). Pore-structure heterogeneity is therefore amplified into flow heterogeneity, and the amplification itself grows with structural complexity. Further details on sample characteristics and the calculation of $CV_r$ and $CV_v$ are provided in Text~S1 in the Supporting Information~S1.

\FloatBarrier

\subsection{Micro-Continuum Modeling}

To provide a precise quantitative characterization of dissolution regimes, we performed micro-continuum simulations on 3D micro-CT images of the three rock samples by solving the steady-state Stokes equation in the pore space using the improved volume of solid (iVOS) method \cite{maes2022improved, soulaine2024micro}, implemented in the open-source solver GeochemFoam \cite{maes2021geochemfoam}. The model solves flow, transport and reaction by volume-averaging velocity, pressure and concentration. CO$_2$-saturated brine at 10\,MPa and 50\,$^\circ$C was injected as the reactive fluid, with all mineral content treated as calcite. $Pe$ and $Da$ are defined as $Pe = v_\mathrm{avg} L_c / D_m$ and $Da = k_c L_c / D_m$, where $v_\mathrm{avg}$ is the mean pore velocity, $L_c$ the characteristic length, $D_m$ the molecular diffusion coefficient and $k_c$ the surface reaction rate constant. Due to the fixed values of $k_c$, $L_c$ and $D_m$ for a given sample, $Da$ remains constant while $Pe$ is varied by adjusting the injection flow rate. All simulations operate firmly in the reaction-limited conditions ($Pe\cdot Da_\mathrm{adv} < 0.05$, Text~S4 in the Supporting Information~S1), so the regime differences observed below cannot be attributed to differences in $Da$. Comprehensive details on the governing equations, parameters and computational setting are provided in Text~S2 in the Supporting Information~S1. For each sample, the reactive transport was simulated across $Pe$ spanning over three orders of magnitude ($Pe = 1$--$2000$) to  enable cross-sample comparison.

\section{Results and Discussion}

\subsection{Time-dependent Porosity–Permeability Coupling as a Determinant of the Power-law Exponent $n$}

Traditionally, the increase in permeability $K$ during dissolution is assumed to have a power-law dependence on porosity $\phi$ \cite{civan2001scale, bernabe2003permeability}:
\begin{equation}
    {K\over K_0}=\left({\phi\over\phi_0}\right)^n
    \label{Eq:K-phi}
\end{equation}
where 0 represents the initial conditions. The exponent $n$ has been estimated in previous work and spans a wide range from 2-40 \cite{hommel2018porosity, carman1997fluid, noiriel2004investigation, menke2015dynamic, menke2016reservoir, menke2017dynamic, menke20184d, al2017reaction, al2018reservoir, al2019pore}. We will demonstrate that assuming a fixed value of $n$ is incorrect, as its effective value changes over time. We define an instantaneous exponent from the derivative of Eq.~(\ref{Eq:K-phi}):
\begin{equation}
    n = {d\ln{K}\over{d\ln{\phi}}}
    \label{Eq:neff}
\end{equation}

\begin{figure}[!ht]
 \centering
 \includegraphics[width=\textwidth]{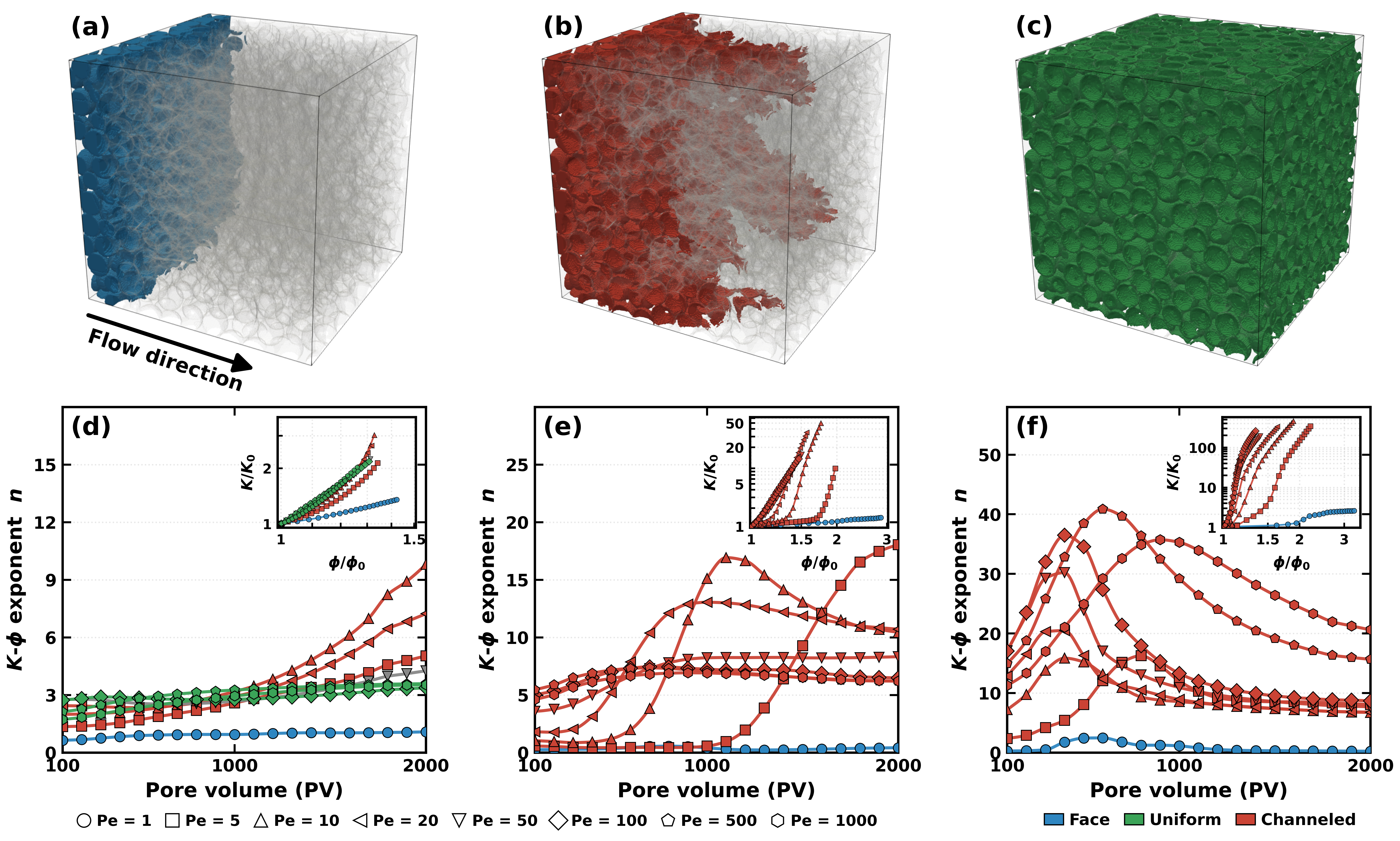}
 \caption{(a--c) 3D visualizations of the dissolved solid with semi-transparent pore space in the beadpack at $\phi{=}1.2\phi_0$ for $Pe{=}1, 10, 1000$, illustrating, respectively, face, channeled, and uniform dissolution. (d--f) Evolution of the instantaneous permeability-porosity exponent $n$ under various $Pe$ for (d) the beadpack, (e) Ketton, and (f) Estaillades. Curves are colored by dissolution regime: face (blue), uniform (green) and channeled (red). The insets show the corresponding normalized porosity--permeability ($(K/K_0)-(\phi/\phi_0)$) relationships.}
 \label{fig:fig2}
\end{figure}

Figure~\ref{fig:fig2}a--c illustrates the three dissolution regimes: face, channeled and uniform by showing dissolved solid for the beadpack at $Pe$ = 1, 10, and 1000 respectively. Figure~\ref{fig:fig2}d--f shows the evolution of $n$, Eq.~(\ref{Eq:neff}), with pore-volume injected (PV) for the three samples for various $Pe$. At the lowest flow rate ($Pe = 1$), all three samples exhibit face dissolution, with a similar, slowly rising $n$ trajectory in which weak advection coupled with diffusion drives complete consumption of the acid near the inlet. The average values for $n$ are 0.94, 0.35 and 0.81 for the beadpack, Ketton and Estaillades, respectively. In the limit of high $Pe$ for the beadpack, we have uniform dissolution, which is also characterized by nearly constant $n$, with average $n = 2.95$ at $Pe = 100$. Face dissolution at low $Pe$ and uniform dissolution at high $Pe$ are the two classical limits, easily distinguished by $Pe$ alone. At intermediate $Pe$, the trajectories diverge. In the beadpack (Figure~\ref{fig:fig2}d), $n$ rises monotonically and stays bounded below $\sim$10 for every $Pe$, indicating pore enlargement without significant flow focusing or peak formation. In Ketton (Figure~\ref{fig:fig2}e), the trajectory splits into two distinct behaviors.  At intermediate $Pe$, $n$ develops a clear peak as channels transiently emerge, with the peak shifting to later PV as $Pe$ decreases, at higher $Pe$, channels emerge early and $n$ stabilizes on a plateau without subsequent decay. In Estaillades (Figure~\ref{fig:fig2}f), $n$ exhibits a sharp single peak followed by a steep decay across nearly all $Pe$, where the transient channel formation is followed by progressive channel widening. The trajectory alone discriminates the three samples, indicating that dissolution behavior is a dynamic trajectory rather than a static classification. The non-monotonic evolution of pore-scale heterogeneity drives the corresponding modulation of the exponent $n$, and demonstrates that no single power-law exponent in Eq.~(\ref{Eq:K-phi}) can describe the full $K$--$\phi$ trajectory once channeled dissolution sets in.

\FloatBarrier

\subsection{Feasible Dissolution Regimes as a Function of Heterogeneity}

\begin{figure}[!ht]
 \centering
 \includegraphics[width=\textwidth]{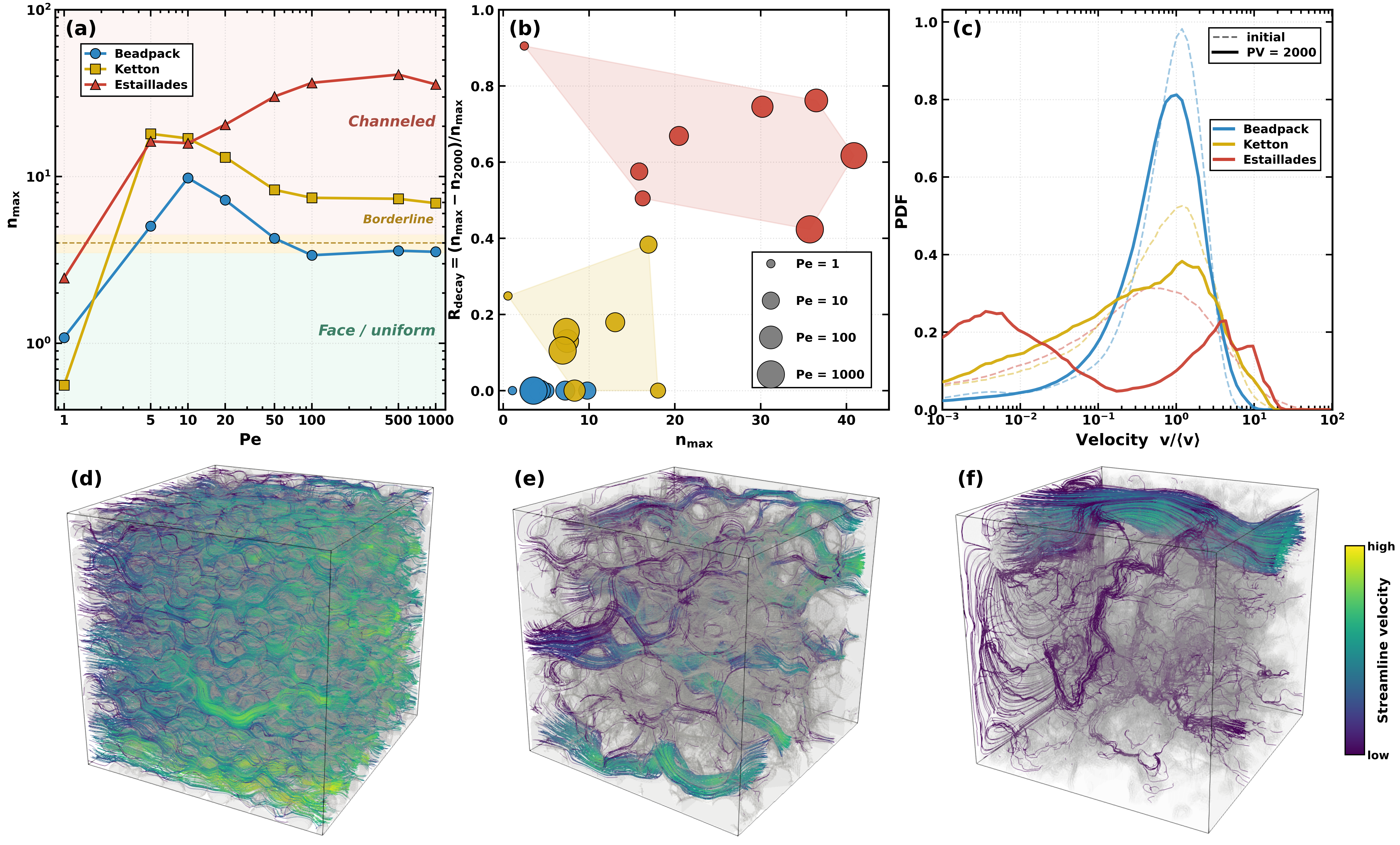}
 \caption{(a) The maximum exponent, $n_{\mathrm{max}}$ as a function of $Pe$ for the three rock samples. The dashed line at $n_\mathrm{max}{=}4$ marks the channeled dissolution threshold, the shaded band marks the borderline zone ($3.5{\leq}n_\mathrm{max}{\leq}4.5$). (b) Channel decay ratio $R_\mathrm{decay}{=}(n_\mathrm{max}{-}n_{2000})/n_\mathrm{max}$ versus $n_\mathrm{max}$, where $n_{2000}$ is the exponent at 2000 PV, with marker size scaled to $\log(Pe)$. Shaded convex hulls enclose the data range of each rock. (c) Velocity PDFs at $Pe{=}100$ at the initial state (dashed lines, PV\,${=}0$) and final dissolution (solid lines, PV\,${=}2000$). (d--f) Corresponding 3D streamlines within the semi-transparent mineral at $Pe{=}100$ and PV\,${=}2000$ for (d) the beadpack, (e) Ketton, and (f) Estaillades, colored by velocity magnitude.}
 \label{fig:fig3}
\end{figure}

Figure~\ref{fig:fig3}a summarizes the maximum exponent $n_\mathrm{max}$ as a function of $Pe$ for the three samples. The three samples occupy distinct ranges of $n_\mathrm{max}$ that follow $CV_v$: the beadpack ($CV_v = 0.91$) stays at $n_\mathrm{max} \leq 10$ across the full $Pe$ range; Ketton ($CV_v = 1.61$) occupies an intermediate band; Estaillades ($CV_v = 2.62$) reaches $n_\mathrm{max} \approx 40$ near $Pe = 500$, an order of magnitude above the homogeneous reference. The height of $n_\mathrm{max}$ reflects the instantaneous flow contrast once the channel has emerged, while the subsequent decay (Figure~\ref{fig:fig2}) tracks the how uniformly the flow reorganizes thereafter. both grow with $CV_v$, which therefore controls both the onset of channel emergence and the flow reorganization that follows.

To distinguish these trajectories into different regimes, we use $n_\mathrm{max}$ as the diagnostic indicator. The baseline Carman--Kozeny relation ($K \propto \phi^3$) gives $n = 3$, while the full relation $K \propto \phi^3/(1-\phi)^2$ \cite{carman1997fluid} yields $n \approx 4.1$ for the beadpack. The simulated beadpack asymptote of $n \approx 3.4$--$3.6$ at high $Pe$ falls between these two limits, justifying $n_\mathrm{max} = 4$ with a $\pm\,0.5$ buffer as the channeled dissolution threshold (Text~S5 in the Supporting Information~S1). Trajectories with $n_\mathrm{max} < 3.5$ are classified as face/uniform and those with $n_\mathrm{max} \geq 4.5$ as channeled dissolution, with the intermediate window $3.5 \leq n_\mathrm{max} \leq 4.5$ marking a borderline zone, as color-coded in the $n$ trajectories of Figure~\ref{fig:fig2} (see also the beadpack velocity PDFs in Figure~S2). Whether a rock reaches uniform dissolution at any $Pe$ depends on its initial pore-scale heterogeneity, the homogeneous beadpack remains nearly uniform across most of the $Pe$ range, whereas the heterogeneous Ketton and Estaillades enter channeled dissolution at $Pe \geq 5$ and remain channeled at higher $Pe$. Additionally, the $Pe$ at which $n_\mathrm{max}$ reaches a maximum also shifts with $CV_v$. The beadpack and Ketton both reach a maximum near $Pe \approx 10$. Estaillades, in contrast, reaches its maximum $n$ near $Pe \approx 500$, nearly two orders of magnitude higher, indicating that in strongly heterogeneous media the structural heterogeneity selects the transport condition under which preferential channels are most efficiently exploited.

Figure~\ref{fig:fig3}b plots the channel decay ratio $R_\mathrm{decay} = (n_\mathrm{max} - n_{2000})/n_\mathrm{max}$ against $n_\mathrm{max}$, where $n_{2000}$ is the exponent at 2000 PV. The three samples occupy distinct regions. In the beadpack at all $Pe$ and Ketton at high $Pe$, $R_\mathrm{decay}$ is near zero: channels remain stable after they emerge. In Estaillades at moderate to high $Pe$, $R_\mathrm{decay} = 0.5$--$0.9$: a single channel captures most of the flow, leaving the rest of the pore space nearly stagnant. $R_\mathrm{decay}$ adds a dynamic dimension to the $n_\mathrm{max}$ partition, with the ordering again following $CV_v$. Figure~\ref{fig:fig3}c shows the velocity PDFs at $Pe = 100$ and $\mathrm{PV} = 2000$, the beadpack PDF remains narrow and unimodal around $v/\langle v\rangle = 1$, Ketton broadens asymmetrically toward lower velocities, and Estaillades develops a bimodal structure with stagnant ($v/\langle v\rangle \sim 10^{-3}$) and high-velocity ($v/\langle v\rangle \sim 5$) modes. The 3D streamlines (Figure~\ref{fig:fig3}d--f) show the geometry behind these statistical signatures, uniform flow in the beadpack, competing paths in Ketton, and a single channel through a stagnant matrix in Estaillades. These observations show that the velocity distributions and $CV_v$ determine the dissolution regime each rock follows.

\FloatBarrier
\subsection{Channeled Dissolution and Effective Reaction Rate}

\begin{figure}[!ht]
 \centering
 \includegraphics[width=\textwidth]{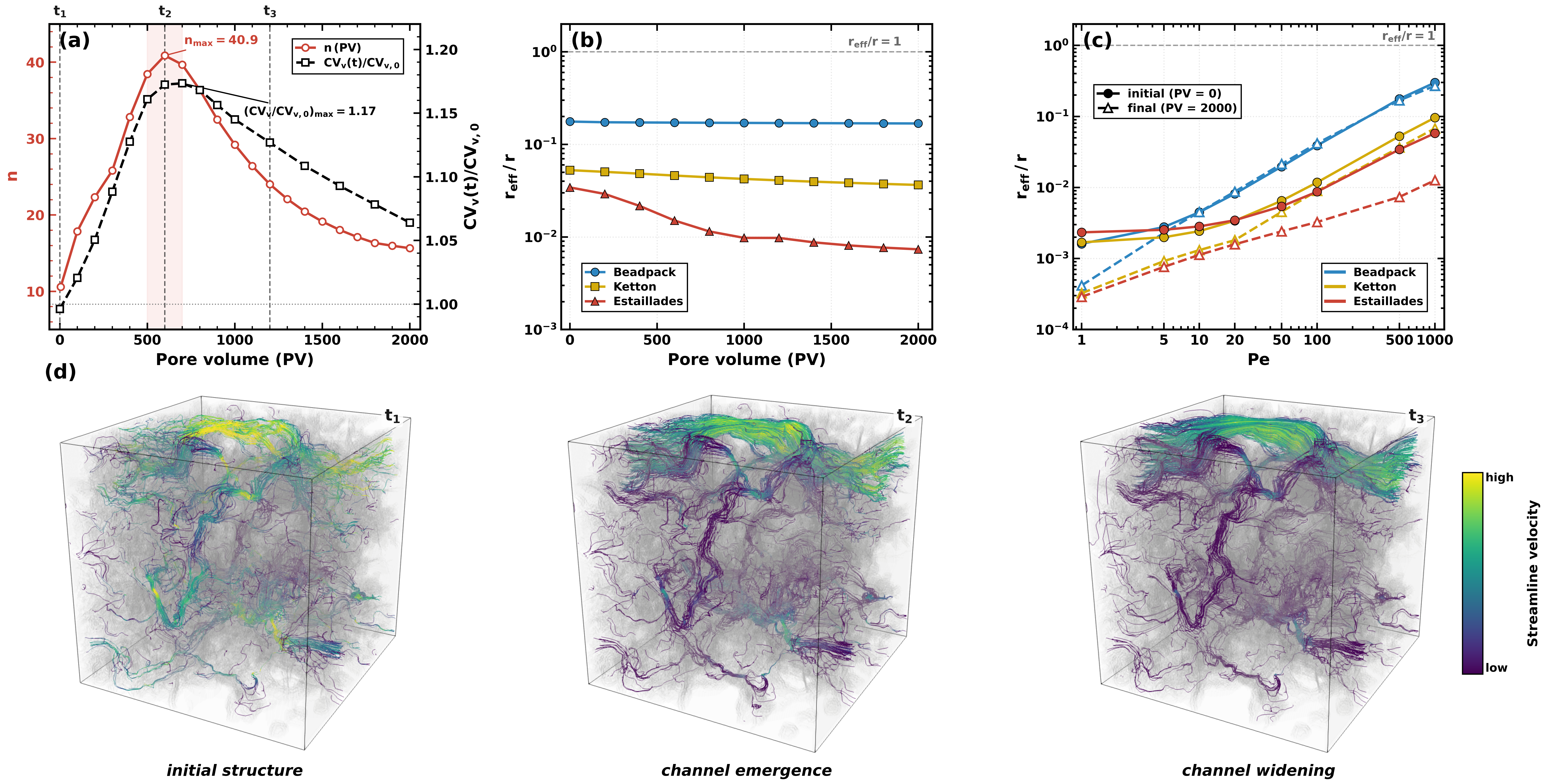}
 \caption{(a) Coevolution of the exponent $n$ (left axis) and the normalized flow heterogeneity $CV_v(t)/CV_{v,0}$ (right axis) versus PV. Both quantities reach a maximum simultaneously, defining a transient channel emergence window. (b) The normalized effective reaction rate $r_\mathrm{eff}/r$ as a function of PV at $Pe{=}500$ for the three samples. The horizontal dashed line marks the uniform dissolution prediction $r_\mathrm{eff}/r{=}1$. (c) $r_\mathrm{eff}/r$ as a function of $Pe$ at the initial state (solid lines) and final dissolution (dashed lines). (d) 3D streamline behaviors at $t_1$ (PV\,${=}0$, initial), $t_2$ (PV\,${=}600$, channel emergence), and $t_3$ (PV\,${=}1200$, channel widening), colored by velocity magnitude.}
 \label{fig:fig4}
\end{figure}

The $CV_v$ identifies which dissolution regimes are feasible for each rock. We now examine the dynamic event of channeled dissolution and its effect on the macroscopic dissolution rate. We focus on Estaillades at $Pe = 500$, the case with the maximum exponent ($n_\mathrm{max} = 40.9$; Figure~\ref{fig:fig3}a). Figure~\ref{fig:fig4}a tracks the joint evolution of the exponent $n(\mathrm{PV})$ and the normalized flow heterogeneity $CV_v(t)/CV_v$, as a function of PV. Figure~\ref{fig:fig4}d visualizes the corresponding 3D streamline behaviors at three characteristic stages: $t_1$ (PV\,$=0$, initial), $t_2$ (PV\,$=600$, channel emergence), and $t_3$ (PV\,$=1200$, channel widening), and the corresponding velocity PDF evolution is shown in Figure~S3. From PV\,$=0$, $n$ and $CV_v(t)/CV_v$ both reach a maximum within a localized PV window centered at PV\,$\approx 600$ with $n_\mathrm{max} = 40.9$ and $CV_{v,\mathrm{max}}/CV_v = 1.17$. Beyond this window both quantities decrease with continued injection. This indicates that the transition between dissolution regimes is a localized dynamic event rather than a gradual reclassification. $n$ and $CV_v(t)$ reach their maximum simultaneously across all six Estaillades channeled dissolution cases ($Pe = 5$--$1000$), with the two maximum falling within a $\pm\,100$ PV envelope (Figure~S4).

The 3D streamlines (Figure~\ref{fig:fig4}d) and velocity PDF (Figure~S3) show that this window marks a transition in both flow topology and velocity distribution. At $t_1$ ($\mathrm{PV} = 0$), the velocity PDF is unimodal near $v/\langle v\rangle \approx 0.4$ and the streamlines are widely distributed, reflecting the high $CV_v$ of Estaillades. Within the channel emergence window, at $t_2$ ($\mathrm{PV} = 600$), the PDF becomes bimodal with a low-velocity peak near $v/\langle v\rangle \sim 10^{-2}$ and a high-velocity peak near $v/\langle v\rangle \sim 5$, and the streamlines correspondingly collapse onto a single preferential channel. Beyond the window, at $t_3$ ($\mathrm{PV} = 1200$), the PDF remains bimodal and the channel keeps widening, while the rest of the pore space stays stagnant. The unimodal to bimodal PDF transition and the emergence of a single preferential flow path thus describe the same reorganization, localized within the PV window in which $n$ and $CV_v(t)$ reach a maximum.

To assess the impact of heterogeneity and transport conditions on reaction rate, we evaluated the effective reaction rate $r_\mathrm{eff}$ (Text~S6 in the Supporting Information~S1) and compare it with the intrinsic calcite surface reaction rate $r = 8.1 \times 10^{-4}$ mol m$^{-2}$ s$^{-1}$. Figure~\ref{fig:fig4}b shows $r_\mathrm{eff}/r$ as a function of $\mathrm{PV}$ at $Pe = 500$, the trajectories diverge sharply across $CV_v$, with the beadpack pseudo-steady, Ketton decaying mildly, and Estaillades undergoing strong monotonic decay, indicating that mass transfer limitations grow with flow heterogeneity. Figure~\ref{fig:fig4}c compares $r_\mathrm{eff}/r$ across the full $Pe$ range at PV\,$=0$ (initial) and PV\,$=2000$ (after dissolution). All values lie one to three orders of magnitude below the uniform dissolution prediction $r_\mathrm{eff}/r = 1$, and the gap between the initial and final curves widens with $CV_v$, isolating the additional suppression introduced by flow field reorganization driven by dissolution from the intrinsic geometric suppression of the initial pore structure. Only the homogeneous beadpack approaches the uniform dissolution limit. In the heterogeneous Ketton and Estaillades, channeled dissolution reduces the effective reaction rate below the value predicted assuming uniform consumption, confirming that channeled dissolution slows mineral dissolution in heterogeneous porous media.

\FloatBarrier
\section{Conclusions}
We have shown that the dissolution regime in porous rocks is best understood as a dynamic trajectory rather than a fixed label in the P\'eclet--Damk\"ohler ($Pe$--$Da$) plane. Across three samples spanning a wide range of pore-scale heterogeneity, the initial flow heterogeneity captured by the coefficient of variation $CV_v$ governs both the regime each rock can develop and the optimal $Pe$ at which channeled dissolution produces the steepest rise of permeability with porosity, charaterised by the instantaneous power-law exponent $n$.

The homogeneous beadpack sustains uniform dissolution with $n_{\max} \leq 10$ across most of the $Pe$ range, while the heterogeneous carbonates are driven into channeled dissolution regardless of $Pe$, with $n_{\max}$ an order of magnitude higher and the optimal $Pe$ shifting from $\approx 10$ in Ketton to $\approx 500$ in Estaillades. Channel emergence is a dynamic event that simultaneously reorganizes the $K$--$\phi$ coupling and the velocity PDF, suppressing the effective reaction rate by one to three orders of magnitude, identifying the dynamic reorganization of flow, rather than the intrinsic surface kinetics, as the dominant control on the macroscopic dissolution rate of heterogeneous rocks.

These results were obtained in the reaction-limited conditions for three monomineralic samples. Extending the framework to diffusion-limited conditions and multimineral rocks is a topic for future work. The findings imply that reservoir-scale models of CO$_2$ storage, geothermal extraction and karst evolution need to move beyond static $Pe$--$Da$ classifications toward formulations that explicitly resolve the coevolution of pore-scale heterogeneity and flow.

\section*{Open Research Section}
Micro-CT images are publicly available at the Digital Porous Media Portal (beadpack: \url{doi.org/10.17612/P73W2C}; Estaillades: \url{doi.org/10.17612/P78G69}) and Zenodo (Ketton: \url{zenodo.org/records/17856861}). Simulations were performed with the open-source solvers GeochemFoam (\url{github.com/GeoChemFoam}) and pnextract (\url{github.com/ImperialCollegeLondon/porescale}). Result data are publicly available on Zenodo \cite{wang2026data}.

\section*{Acknowledgments}
Jinlei Wang gratefully acknowledges financial support from the China Scholarship Council. Yongfei Yang acknowledges the National Natural Science Foundation of China (Nos.\ U23A20595 and 52288101). Computational resources were provided by Imperial College London and the SuperComputing Network (SCNet).


\bibliographystyle{apacite}
\bibliography{reference}

\end{document}